\documentclass[useAMS,usenatbib]{mn2e}

\usepackage{amssymb}
\usepackage{times}
\usepackage{graphicx}

\newcommand{\beq}{\begin{equation}}
\newcommand{\eeq}{\end{equation}}

\newcommand{\bfd}{\mathbf{d}}

\newcommand{\bfP}{\mathbf{P}}

\def\gs{\mathrel{\lower0.6ex\hbox{$\buildrel {\textstyle >}\over{\scriptstyle \sim}$}}}
\def\ls{\mathrel{\lower0.6ex\hbox{$\buildrel {\textstyle <}\over{\scriptstyle \sim}$}}}
\newcommand{\simgt}{\lower.5ex\hbox{$\; \buildrel > \over \sim \;$}}
\newcommand{\simlt}{\lower.5ex\hbox{$\; \buildrel < \over \sim \;$}}

\newcommand{\aap}{A\&A}
\newcommand{\apj}{ApJ}
\newcommand{\apjl}{ApJ}
\newcommand{\apjs}{ApJS}
\newcommand{\aj}{AJ}
\newcommand{\pasj}{PASJ}

\newcommand{\mnras}{MNRAS}

\begin{document}

\title[Concentrations and orientations of the $z > 0.5$ MACS sample]{Triaxial strong-lensing analysis of the $z > 0.5$ MACS clusters: the mass-concentration relation}
\author[M. Sereno, A. Zitrin]{
Mauro Sereno$^{1,2}$\thanks{E-mail: mauro.sereno@polito.it (MS)} and Adi Zitrin$^{3}$
\\
$^1$Dipartimento di Fisica, Politecnico di Torino, corso Duca degli Abruzzi 24, 10129 Torino, Italia\\
$^2$INFN, Sezione di Torino, via Pietro Giuria 1, 10125, Torino, Italia\\
$^3$The School of Physics and Astronomy, the Raymond and Beverly Sackler Faculty of Exact Sciences, Tel Aviv University, Tel Aviv 69978, Israel
}


\maketitle

\begin{abstract}
The high concentrations derived for several strong-lensing clusters present a major inconsistency between theoretical $\Lambda$CDM expectations and measurements. Triaxiality and orientation biases might be at the origin of this disagreement, as clusters elongated along the line-of-sight would have a relatively higher projected mass density, boosting the resulting lensing properties. Analyses of statistical samples can probe further these effects and crucially reduce biases. In this work we perform a fully triaxial strong-lensing analysis of the 12 MACS clusters at $z > 0.5$, a complete X-ray selected sample, and fully account for the impact of the intrinsic 3D shapes on their strong lensing properties. We first construct strong-lensing mass models for each cluster based on multiple-images, and fit projected ellipsoidal Navarro-Frenk-White halos with arbitrary orientations to each mass distribution. We then invert the measured surface mass densities using Bayesian statistics. Although the Einstein radii of this sample are significantly larger than predicted by $\Lambda$CDM, here we find that the mass-concentration relation is in full agreement with results from $N$-body simulations. The $z > 0.5$ MACS clusters suffer from a moderate form of orientation bias as may be expected for X-ray selected samples. Being mostly unrelaxed, at a relatively high redshift, with high X-ray luminosity and noticeable substructures, these clusters may lie outside the standard concentration-Einstein radius relation. Our results remark the importance of triaxiality and properly selected samples for understanding galaxy clusters properties, and suggest that higher-$z$, unrelaxed low-concentration clusters form a different class of prominent strong gravitational lenses. Arc redshift confirmation and weak lensing data in the outer region are needed to further refine our analysis.

\end{abstract}


\section{Introduction}\label{intro}

The hierarchical cold dark matter model with a cosmological constant ($\Lambda$CDM) is highly successful in explaining many features of galaxy clusters. The universal Navarro-Freank-White (NFW) density profile \citep{nfw96,nav+al97} reproduces well their density profile over most radii, but the actual mass-concentration relation, $c(M)$, is still debated \citep{co+na07,bro+al08}.

The concentration measures the halo central density relative to outer parts, and is known to correlate reversely with the virial mass and redshift, so that lower mass and lower redshift clusters would show higher concentrations \citep{bul+al01,duf+al08}. However, recent high-resolution simulations show different trends than have been expected before, with a turn-around and decrease in concentrations towards very high virial-mass clusters \citep{pra+al11}. The cluster baryonic content is also expected to play a fundamental role in shaping the cluster overall density profile, although the baryonic physics is yet to be fully accounted for in related simulations, so that the explicit effects of cooling, feedback, and baryonic-DM interplay on the mass profile and concentration are still ambiguous \citep[e.g.,][]{gne+al04,Roz+al08,Duffy10}. Cluster observations have also yet to firmly assess the properties of the $c(M)$ relation, due to the small numbers of clusters analysed to date. Extensive lensing analyses done in recent years \citep[e.g.,][]{ogu+al09,zit+al09b,zit+al10,oka+al10,ric+al10,ume+al11a}, and ongoing cluster lensing surveys (e.g., CLASH, \citealt{Post+al11CLASH}) should help in characterising further the observed relation.

Similar discrepancies in lensing clusters between the observed properties and those predicted by $\Lambda$CDM have been reported, such as the arc-abundance problem \citep{Bar+al98}, and the detection of several extremely large Einstein radii \citep{br+ba08,og+bl09,zit+al11,zit+al11b}. In fact, these discrepancies are highly related to the over-concentration problem, as there is a clear connection between the concentration and the Einstein radius \cite[e.g.,][]{sa+re08}, or the lensing cross-section and the efficiency for generating giant arcs \citep{hen+al07}. In addition, the existence of high-redshift massive clusters adds to this tension, and may imply a non-Gaussian distribution of massive perturbations \citep{Chon+Silk2011,D'Al+Nat11}, or an earlier growth of structure than entailed by $\Lambda$CDM, so that clusters, sitting atop the hierarchical build-up, would have had more time to bind together and concentrate, increasing their inner mass and lensing properties.

The disagreement between theory and observation might be explained by some orientation and shape biases \citep{ogu+al05,se+um11}. In clusters that are elongated along the line-of-sight the projected matter density is relatively higher, and the observed lensing properties are boosted \citep{hen+al07}. Correspondingly, neglecting halo triaxiality can lead to over- and under-estimates of up to 50 per cent and a factor of 2 in halo mass and concentration, respectively \citep{cor+al09}. Assuming spherical symmetry also causes underestimation of statistical uncertainties \citep{cor+al09,se+um11}. \citet{ser+al10} investigated a sample of 10 strong lensing (SL) clusters considering which intrinsic shape and orientation the lensing halos should have to account for both theoretical predictions and observations. They found that nearly one half of the clusters seemed to be composed of outliers of the mass-concentration relation, whereas the second half supported expectations of $N$-body simulations which prefer mildly triaxial lensing clusters with a strong orientation bias. Recently, \citet{se+um11} performed a full triaxial weak and strong lensing analysis of A1689, a cluster usually seen as strongly over-concentrated. They found the halo to be slightly over-concentrated but still consistent with theoretical predictions. They also found some evidence for a mildly triaxial lens with the major axis orientated along the line of sight.

Here, we try to assess the impact of halo triaxiality and orientation on the mass-concentration relation of a properly defined statistical sample. Recently, \citet{zit+al11} presented a detailed strong lensing analysis of a complete sample of X-ray selected clusters, the MACS high-redshift cluster sample \citep{ebe+al07}. This sample comprises 12 very luminous X-ray galaxy clusters at $z > 0.5$. \citet{zit+al11} reported a difference of $\sim 40$ per cent between the observed and the theoretical distributions of Einstein radii in standard $\Lambda$CDM based on analytic models of galaxy clusters. \citet{men+al11} extracted clusters from the MareNostrum Universe to build up mock catalogs of clusters selected through their X-ray flux according to the criteria applied to the $z>0.5$ MACS sample. Their simulated distribution of the Einstein ring sizes is also lower by $\sim 25$ per cent than that found in observations. They also found that orientation biases together with triaxiality issues still affect the lenses with the largest cross sections, and that the concentrations of the individual MACS clusters inferred from the lensing analysis might be up to a factor of $\sim 2$ larger than expected. They suggested that for a significant fraction of  $\sim 20$ per cent of the clusters in the MACS sample, the lensing-derived concentrations should be higher than expected by more than $\gs 40$ per cent.

It should also be noted that this high-$z$ MACS sample appears to consist of mostly unrelaxed clusters. The high X-ray luminosity, along with the relatively high redshift and recent Sunyaev-Zeldovich (SZ) effect images (unpublished; private communication), are in support of this claim, in addition to noticeable substructure seen in the optical and the aforementioned lensing analysis \citep{zit+al11}. The inner mass distribution of (most of) these clusters seem correspondingly rather widely-distributed, which may account for their extensive lensing properties and large Einstein radii. In fact, it is known that for e.g., relaxed, lower-$z$ clusters, the Einstein radius correlates with the concentration \citep{sa+re08}. More concentrated clusters have more mass in the middle, thus boosting their Einstein radius. For higher-$z$, unrelaxed clusters, large Einstein radii may form due to a widely-distributed inner mass distribution, so that the critical curves of the smaller substructures near the core, merge together to form a much larger Einstein radius curve \citep[e.g.,][]{Tor+al04,Fed+al06,zit+al09}. If this is indeed the case, for such a sample low concentrations and relatively shallow inner profiles should be expected \citep[e.g.,][]{net+al07}, suggesting that unrelaxed (higher-$z$) clusters form a distinct class of impressive strong-lenses.

In order to probe the 3D intrinsic shapes of the clusters, we apply here the 3D strong-lensing analysis method of \citet{se+um11} apart from a main difference. In fitting the projected surface density, we adopt a non-parametric technique, as implemented in the PixeLens software \citep{sa+wi04}, instead of a parametric modelling (though initially the multiple-images were found using a parametric method, see below). As in \citet{se+um11}, we use Bayesian statistics to invert the measured projected surface mass-densities. An additional important difference is that here we use only strong-lensing data, without implementing full range weak-lensing data, as no such analyses are currently available for this sample. In that context, upcoming weak-lensing analyses for these clusters (e.g., as part of the CLASH program, \citealt{Post+al11CLASH}) will be important to further establish the results of our work.

\citet{zit+al11}, thanks to their minimalistic parametric approach to lensing and the low number of free parameters in their modelling, were able to find many multiple images across the $z > 0.5$ MACS cluster fields, which we use here as input. It should be noted, that although the mass distributions of these clusters (and resulting critical curves and Einstein radii) were credibly constrained therein, due to lack of redshift information for most of the lensed features, only broad constraints were put on the (projected) mass profiles of the different clusters, following the models best reproducing the different multiple images \citep[see][]{zit+al11}. Here we use the \citet{zit+al11} best-model predicted redshifts for the different arcs, in order to maintain consistency with their input mass distributions. These redshift estimates were found to be usually very accurate, in following spectroscopic (e.g., \citealt{Smi+al09}) and photometric (the CLASH program, in preparation) measurements of some of these clusters.

Since our strong lensing analysis with PixeLens exploited the image positions and source redshifts inferred in \citet{zit+al11}, our modelling is not alternative. The main advantage of re-analysing the strong lensing features with an independent non-parametric technique is that PixeLens allows us to fully explore degeneracies in the projected mass density. We can then reproduce the best-fitting models of \citet{zit+al11} and obtain a reliable estimate of the covariance matrices, that could be then used for optimising the deprojection procedure.

The paper is organised as follows: in \S~\ref{sec_nfw} we detail the basics of a triaxial NFW description, whereas in \S~\ref{sec_theory} we discuss the theoretical predictions for the results. In \S~\ref{sec_sample} the 12 high-$z$ MACS sample is described, and in \S~\ref{sec_stron} the strong-lensing analysis procedure is laid out, followed by the deprojection procedure in \S~\ref{sec_depr}. The results are detailed in \S~\ref{sec_results}. In \S~\ref{sec_syst} we review some systematics. Results are concluded in \S~\ref{sec_conc}. Throughout the paper, we assume a flat $\Lambda$CDM cosmology with density parameters $\Omega_\mathrm{M}=0.3$, $\Omega_{\Lambda}=0.7$ and Hubble constant $H_0=100h~\mathrm{km~s}^{-1}\mathrm{Mpc}^{-1}$, $h=0.7$ \citep{kom+al11}.

\section{Basics on triaxial NFW halos}
\label{sec_nfw}

High resolution $N$-body simulations have shown that the density distributions of dark matter halos are successfully described as triaxial ellipsoidal NFW density profiles with aligned, concentric axes \citep{nfw96,nav+al97,ji+su02}. The 3D distribution follows
\begin{equation}
\label{nfw1}
	\rho_\mathrm{NFW}=\frac{\rho_\mathrm{s}}{(\zeta/r_\mathrm{s})(1+\zeta/r_\mathrm{s})^2},
\end{equation}
where $\zeta$ is the ellipsoidal radius. The shape is determined by the axial ratios. The minor (intermediate) to major axial ratio is denoted as $q_1$ ($q_2$) with $0< q_1 \le q_2\le 1$; we also use the inverse ratios, $0< e_i =1/q_i \ge 1$.

Three Euler's angles, $\theta, \varphi$ and $\psi$ determine the orientation of the halo. $\theta$ and $\varphi$ fix the orientation of the line of sight in the intrinsic system.  In particular, $\theta$ quantifies the inclination of the major axis with respect to the line of sight. The third angle $\psi$ determines the orientation of the cluster in the plane of the sky.

The radius $r_{200}$ for an NFW spheroid can be defined such that the mean density contained within an ellipsoid of semi-major axis $r_{200}$ is $\Delta= 200$ times the critical density at the halo redshift, $\rho_\mathrm{cr}$ \citep{cor+al09,ser+al10,se+um11}; the corresponding concentration is $c_{200} \equiv r_{200}/ r_\mathrm{s}$. $M_{200}$ is the mass within the ellipsoid of semi-major axis $r_{200}$, $M_{200}=(800\pi/3)q_1q_2 r_{200}^3 \rho_\mathrm{cr}$.

The ellipsoidal 3D NFW halo projects on an elliptical 2D profile. The projected surface density profile has the same functional form of a spherically symmetric profile \citep{sta77,ser07,ser+al10b}. The convergence $\kappa$, i.e., the surface mass density in units of the critical surface mass density for lensing, $\Sigma_\mathrm{cr}=(c^2\,D_\mathrm{s})/(4\pi G\,D_\mathrm{d}\,D_\mathrm{ds})$, where $D_\mathrm{s}$, $D_\mathrm{d}$ and $D_\mathrm{ds}$ are the source, the lens and the lens-source angular diameter distances respectively, for an ellipsoidal NFW halo is
\beq
\kappa_\mathrm{NFW}(x)=2 \kappa_\mathrm{s}\frac{1}{1-x^2}\left[ \frac{1}{\sqrt{1-x^2}} \mathrm{arccosh}\left(\frac{1}{x}\right) -1\right];
\eeq
$x$ is the dimensionless elliptical radius,
\beq
x \equiv \xi /r_\mathrm{sP}, \ \ \xi= [x_1^2 +x_2^2/(1-\epsilon)^2)]^{1/2},
\eeq
where $\epsilon$ is the ellipticity and $x_1$ and $x_2$ are the abscissa and the ordinate in the plane of the sky oriented along the the ellipse axes, respectively.

The ellipticity and the orientation of the projected ellipses depend only on the intrinsic geometry and orientation of the system. The axial ratio of the major to the minor axis of the observed projected isophotes, $e_\mathrm{p }= (1-\epsilon)^{-1}$, can be written as \citep{bin80},
\begin{equation}
\label{eq:tri4e}
e_{\rm p}= \sqrt{ \frac{j+l + \sqrt{(j-l)^2+4 k^2 } }{j+l -\sqrt{(j-l)^2+4 k^2 }} },
\end{equation}
where  $j, k$ and $l$ are defined as
\begin{eqnarray}
j & = &  e_1^2 e_2^2 \sin^2 \theta + e_1^2 \cos^2 \theta \cos^2 \varphi   +  e_2^2 \cos^2 \theta \sin^2 \varphi  ,  \label{eq:tri4a} \\
k & = &  (e_1^2 - e_2^2) \sin \varphi \cos \varphi  \cos \theta   ,  \label{eq:tri4b}  \\
l & = &  e_1^2 \sin^2 \varphi + e_2^2 \cos^2 \varphi . \label{eq:tri4c}
\end{eqnarray}

The angle $\psi$ is strictly connected to the orientation angle in the plane of the sky. When the coordinate axes are aligned with the ellipse axes, $\psi$ is the angle between the axes of the observed ellipses and the projection onto the sky of the ellipsoid third axis \citep{bin85,ser07},
\begin{equation}
\label{eq:tri4f}
\psi = \frac{1}{2} \arctan \left[\frac{2 k}{j-l} \right].
\end{equation}

The strength of the lens $\kappa_\mathrm{s}$ and of the projected length scale $r_\mathrm{sP}$ are directly related to the intrinsic parameters $M_{200}$ and $c_{200}$ and can be inferred by fitting lensing maps \citep{ser+al10b}. The observed scale length $r_\mathrm{sP}$ is the projection on the plane of the sky of the cluster intrinsic length $r_\mathrm{s}$ \citep{sta77,ser07},
\begin{equation}
\label{eq:tri6}
r_\mathrm{sP} = r_{\rm s} \left( \frac{e_\Delta}{\sqrt{f}} \right),
\end{equation}
where $f$ is a function of the cluster shape and orientation,
\begin{equation}
\label{eq:tri3}
f = e_1^2 \sin^2 \theta \sin^2 \varphi + e_2^2 \sin^2 \theta \cos^2 \varphi + \cos^2 \theta ;
\end{equation}
the parameter $e_\Delta$ quantifies the elongation of the triaxial ellipsoid along the line of sight \citep{ser07},
\beq
e_\Delta = \left( \frac{e_\mathrm{P}}{e_1 e_2}\right)^{1/2} f^{3/4};
\eeq
$e_\Delta$ is the ratio between the major axis of the projected ellipse in the plane of the sky and the size of the ellipsoid along the line of sight. It quantifies the elongation of the triaxial ellipsoid along the line of sight. If $e_\Delta < 1$, then the cluster is more elongated along the line of sight than wide in the plane of the sky, i.e., the smaller $e_\Delta$, the larger the elongation along the line of sight.

The lensing strength can be expressed as \citep{ser+al10b},
\beq
\label{nfw1b}
\kappa_\mathrm{s} = \frac{1}{\Sigma_\mathrm{cr} }\frac{\rho_\mathrm{s} r_\mathrm{sP}}{e_\Delta},
\eeq
where as usual $\rho_\mathrm{s} = \delta_c \rho_\mathrm{cr}$,
\beq
\delta_c = \frac{200}{3}\frac{c_{200}}{\ln (1+c_{200})-c_{200}/(1+c_{200})}.
\eeq
The mass $M_{200}$ can be expressed as
\beq
\label{nfw2}
M_{200}= \frac{4\pi}{3}\times 200 \rho_\mathrm{cr} \times (c_{200} r_\mathrm{sP})^3 \frac{1}{e_\mathrm{P}e_\Delta}.
\eeq

\section{Theoretical predictions}\label{sec_theory}

\begin{table*}
\caption{Properties of the $z>0.5$ MACS sample. Data in columns 1-8 are based on \citet{ebe+al07}. \emph{Column 1:} cluster name in the MACS survey; \emph{Columns 2 \& 3:} The RA and Declination of the X-ray centroids (as determined from Chandra ACIS-I data); \emph{Column 4:} redshifts; \emph{Column 5:} velocity dispersions, in km s$^{-1}$; \emph{Column 6:} Chandra X-ray luminosities in the $0.1$-$2.4$ keV band. These luminosities are quoted within $r_{200}$ and exclude X-ray point sources; \emph{Column 7:} X-ray temperatures, measured from Chandra data within $r_{1000}$, but excluding a central region of 70~kpc radius around the listed X-ray centroid; \emph{Column 8:} \citet{ebe+al07} morphology code - assessed visually based on the appearance of the X-ray contours and the goodness of the optical/X-ray alignment. The assigned codes (from apparently relaxed to extremely disturbed) are 1 (pronounced cool core, perfect alignment of X-ray peak and single cD galaxy), 2 (good optical/X-ray alignment, concentric contours), 3 (nonconcentric contours, obvious small-scale substructure), and 4 (poor optical/X-ray alignment, multiple peaks, no cD galaxy), errors are estimated as less than 1. Note that indeed most clusters do not seem relaxed; \emph{Column 9:} Einstein radius estimation from \citet{zit+al11}, for a source at $z\sim2\pm0.5$. Note, the clusters MACS J0018.5+1626 and MACS J0454.1-0300 are better known from earlier work as Cl0016+1609 and MS~0451.6-0305, respectively. For more information see \citet{ebe+al07}.
}
\label{table:sample}
\begin{center}
\begin{tabular}{|c|c|c|c|c|c|c|c|c|}
  \hline\hline
  MACS & $\alpha$ & $\delta$& $z$ & $\sigma$ & $L_{x, Chandra}$ & $KT$ & M.C.E. & $\theta_\mathrm{E}$\\
  	 &  [J2000.0] & [J2000.0] & & [km~s$^{-1}$] & $[10^{44}\mathrm{erg~s}^{-1}]$ & [keV] &  & $[\arcsec]$\\
  \hline
  J0018.5+1626 & 00 18 33.835 & +16 26 16.64 & 0.545 & $1420^{+140}_{-150}$ & $19.6\pm{0.3}$ & $9.4\pm{1.3}$ & 3 & $24\pm2$\\
  J0025.4-1222 & 00 25 29.381 & -12 22 37.06 & 0.584 & $740^{+90}_{-110}$ & $8.8\pm{0.2}$  & $7.1\pm{0.7}$ & 3 & $30\pm2$\\
  J0257.1-2325 & 02 57 09.151 & -23 26 05.83 & 0.505 & $970^{+200}_{-250}$ & $ 13.7\pm{0.3}$ & $10.5\pm{1.0}$&  2 & $39\pm2$\\
  J0454.1-0300 & 04 54 11.125 & -03 00 53.77 & 0.538 & $1250^{+130}_{-180}$ & $16.8\pm{0.6}$ & $7.5\pm{1.0}$ & 2 & $13^{+3}_{-2}$\\
  J0647.7+7015 & 06 47 50.469 & +70 14 54.95 & 0.591 & $900^{+120}_{-180}$ & $15.9\pm{0.4}$ & $11.5\pm{1.0}$&  2 & $28\pm2$\\
  J0717.5+3745 & 07 17 30.927 & +37 45 29.74 & 0.546 & $1660^{+120 }_{-130}$ & $24.6\pm{0.3}$ & $11.6\pm{0.5}$&  4 & $55\pm3$\\
  J0744.8+3927 & 07 44 52.470 & +39 27 27.34 & 0.698 & $1110^{+130 }_{-150}$ & $22.9\pm{0.6}$ & $8.1\pm{0.6}$&  2 & $31\pm2$\\
  J0911.2+1746 & 09 11 11.277 & +17 46 31.94 & 0.505 & $1150^{+260 }_{-340}$ & $7.8\pm{0.3}$ & $8.8\pm{0.7}$&  4 & $11^{+3}_{-1}$\\
  J1149.5+2223 & 11 49 35.093 & +22 24 10.94 & 0.544 & $1840^{+120 }_{-170}$ & $17.6\pm{0.4}$ & $9.1\pm{0.7}$ & 4 & $27\pm3$\\
  J1423.8+2404 & 14 23 47.663 & +24 04 40.14 & 0.543 & $1300^{+120}_{-170}$ & $16.5\pm{0.7}$ & $7.0\pm{0.8}$ & 1 & $20\pm2$\\
  J2129.4-0741 & 21 29 26.214 & -07 41 26.22 & 0.589 & $1400^{+170}_{-200}$ & $15.7\pm{0.4}$ & $8.1\pm{0.7}$ & 3 & $37\pm2$\\
  J2214.9-1359 & 22 14 57.415 & -14 00 10.78 & 0.503 & $1300^{+90}_{-100}$ & $14.1\pm{0.3}$ & $8.8\pm{0.7}$ & 2 & $23\pm2$\\
  \hline
\end{tabular}
\end{center}
\end{table*}

$N$-body simulations \citep{net+al07,mac+al08,gao+al08,duf+al08,pra+al11} have provided a quite detailed picture of the expected properties of dark matter halos. Results may depend on parameters such as the overall normalization of the power spectrum, the mass resolution and the simulation volume \citep{pra+al11}. The dependence of halo concentration on mass and redshift can be adequately described by a power law,
\beq
\label{nbod1}
c =A(M/M_\mathrm{pivot})^B(1+z)^C.
\eeq
As reference, we follow \citet{duf+al08}, who used the cosmological parameters from WMAP5 and found $\{A,B,C\}=\{ 5.71 \pm0.12, -0.084 \pm 0.006, -0.47\pm0.04\}$ for a pivotal mass $M_\mathrm{pivot}=2\times10^{12}M_\odot/h$ in the redshift range $0-2$ for their full sample of clusters. The scatter in the concentration about the median $c(M)$ relation is lognormal,
\beq
\label{nbod2}
p(\ln c | M)=\frac{1}{\sigma\sqrt{2\pi}}\exp \left[ -\frac{1}{2} \left(  \frac{\ln c - \ln c(M)}{\sigma}\right) \right],
\eeq
with a dispersion $\sigma (\log_{10} c_{200})=0.15$ for a full sample of clusters \citep{duf+al08}. Recently, \citet{pra+al11} claimed that the dependence of concentration on halo mass and its evolution can be obtained from the root-mean-square fluctuation amplitude of the linear density field. They noticed a flattening and upturn of the relation with increasing mass and estimated concentrations for galaxy clusters substantially larger than results reported in Eq.~(\ref{nbod1}).

The distribution of minor to major axis ratios can be approximated as \citep{ji+su02},
\beq
\label{nbod3}
p(q_1) \propto \exp \left[ -\frac{(q_1-q_\mu/r_{q_1})^2}{2\sigma_\mathrm{s}^2}\right]
\eeq
where $q_\mu=0.54$, $\sigma_\mathrm{s}=0.113$ and
\beq
r_{q_1} = (M_\mathrm{vir}/M_*)^{0.07 \Omega_\mathrm{M}(z)^{0.7}},
\eeq
with $M_*$ the characteristic nonlinear mass at redshift $z$ and $M_\mathrm{vir}$ the virial mass. The conditional probability for $q_2$ goes as
\beq
\label{nbod4}
p(q_1/q_2|q_1)=\frac{3}{2(1-r_\mathrm{min})}\left[ 1-\frac{2q_1/q_2-1-r_\mathrm{min}}{1-r_\mathrm{min}}\right]
\eeq
for $q_1/q_2 \geq r_\mathrm{min} \equiv \max[q_1,0.5]$, whereas is null otherwise. The axial ratios distribution of the lensing population mimics that of the total cluster population \citep{hen+al07}.

For the orientation, we considered a population of randomly oriented clusters with
\beq
\label{flat3}
p(\cos \theta) = 1
\eeq
for $0 \le \cos \theta \le 1$ and
\beq
p(\varphi)=\frac{1}{\pi}
\eeq
for $-\pi/2 \le \varphi \le \pi/2$. The distribution for $\psi$ is the same as that for $\varphi$. For comparison, we also considered a biased population. Semi-analytical \citep{og+bl09} and numerical \citep{hen+al07} investigations showed a large tendency for lensing clusters to be aligned with the line of sight. This condition can be expressed as \citep{cor+al09}
\beq
\label{nbod5}
p(\cos \theta) \propto \exp \left[-\frac{(\cos \theta -1)^2}{2\sigma_\theta^2}\right].
\eeq
A value of $\sigma_\theta=0.115$ can be representative of the orientation bias for massive strong lensing clusters.

\section{The MACS high-redshift cluster sample}\label{sec_sample}

The Massive Cluster Survey (MACS) supplies a complete sample of the very X-ray luminous clusters in the Universe \citep{EbelingMacsCat2001,Ebeling2010MACSALL}. From this catalog, a complete sample of 12 high-$z$ MACS clusters ($z > 0.5$, see Table~\ref{table:sample}) was defined by \cite{ebe+al07}, which have proved very interesting in several follow-up studies including deep X-ray, SZ and HST imaging \citep[see][ and references therein]{zit+al11}. In particular, here we use the multiple-images identified in \cite{zit+al11} as input for our non-parametric mass reconstruction.

Although the MACS clusters are not selected by their lensing features and therefore are not biased in that sense, they are expected to comprise some of the most massive clusters in the Universe, and thus can be expected to produce impressive lenses. As also discussed above in \S~\ref{intro}, by a SL analysis \cite{zit+al11} deduced the distribution of Einstein radii for this sample, which are found to be noticeably larger than expected by $\Lambda$CDM \citep[see also][]{men+al11}, even after taking lensing bias into account.

Here, we deduce the triaxial 3D mass distribution of these clusters, in order to examine whether the $c(M)$ relation of this sample suffers the same discrepancy as found for the Einstein radii. In this context, this high-$z$ MACS sample is of unique interest. Being at a relatively high redshift, and mostly consisting of unrelaxed clusters (see also Table \ref{table:sample}), this sample cannot be expected to follow the usual Einstein radius - concentration relation \citep[e.g.,][]{sa+re08}. Instead, it can shed more light on the relaxation epoch of massive clusters, and show that, if concentrations are low, unrelaxed (and mostly higher-$z$) clusters indeed form a different class of prominent gravitational lenses (than well-relaxed and concentrated lower-$z$ clusters), as previously implied \citep{Tor+al04,Fed+al06,net+al07,zit+al11b}.

\section{Strong lensing analysis}
\label{sec_stron}

\begin{figure}
       \resizebox{\hsize}{!}{\includegraphics{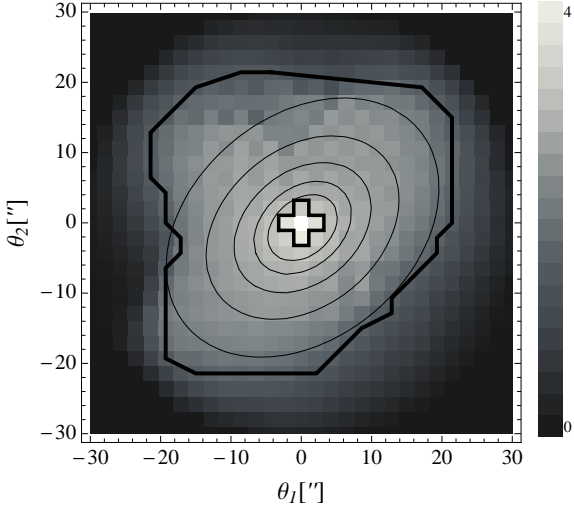}}
       \caption{Pixelated map of the mean convergence of MACSJ0018 as derived with PixeLens. We considered for the analysis only the pixels contained between the inner and the outer thick line. The thin line plot the NFW profile corresponding to the best fit (contours runs from 2.2 to 1.4 in steps of 0.2). North is up.}
	\label{fig_MACSJ0018_density_map}
\end{figure}

\begin{table}
\centering
\caption{Parameters set in the PixeLens analysis. Columns 2 and 3 report the radius of the circular region investigated in pixels and arcseconds, respectively. Column 4 lists the pixel resolution.}
\begin{tabular}[c]{cccc}
        \hline
        \noalign{\smallskip}
        MACS	&	\multicolumn{2}{c}{radius} & resolution     \\
			&	[pixels]		&	$[\arcsec]$	     & [kpc~$h^{-1}$]	\\
	\hline
        J0018 &  14 &  30.0 &  9.6   \\
	J0025 &  14 &  95.0 &  31.4   \\
	J0257 &  14 &  38.0 &  11.7   \\
	J0451 &  14 &  45.0 &  14.3   \\
	J0647 &  14 &  80.0 &  26.6   \\
	J0717 &  15 &  135.9 &  40.5   \\
	J0744 &  14 &  80.0 &  28.6   \\
	J0911 &  14 &  20.0 &  6.1   \\
	J1149 &  15 &  65.0 &  19.3   \\
	J1423 &  15 &  38.0 &  11.3   \\
	J2129 &  14 &  40.0 &  13.3   \\
	J2214 &  14 &  45.0 &  13.8   \\
      \hline
\end{tabular}
\label{tab_PixeLens}
\end{table}

\begin{table}
\centering
\caption{Observed projected NFW parameters of the $z>0.5$ MACS sample. Columns 2, 3, 4, and 5 list the lensing strength $\kappa_\mathrm{s}$, the projected radius $r_\mathrm{sP}$, the ellipticity $\epsilon$ and the orientation angle $\theta_\epsilon$. Central values and dispersions are computed as bi-weight estimators of the marginalised PDFs \citep{bee+al90}. $\theta_\epsilon$ is measured North over East.}
\begin{tabular}[c]{cr@{$\,\pm\,$}lr@{$\,\pm\,$}lr@{$\,\pm\,$}lr@{$\,\pm\,$}l}
        \hline
        \noalign{\smallskip}
        MACS	&	\multicolumn{2}{c}{$\kappa_\mathrm{s}$} &	\multicolumn{2}{c}{$r_\mathrm{sP}$} & \multicolumn{2}{c}{$\epsilon$} & \multicolumn{2}{c}{$\theta_\epsilon$}      \\
			&	\multicolumn{2}{c}{}		&	\multicolumn{2}{c}{[kpc]}	     & \multicolumn{2}{c}{}	&\multicolumn{2}{c}{$[\deg]$}	\\
	\hline
        J0018	&0.29	&0.06	&2200	&1300	&0.32	&0.12	&-50	&15	\\
        J0025	&0.26	&0.04	&8100	&4500	&0.21	&0.09	&-25	&24	\\
        J0257	&0.37	&0.06	&1200	&500	&0.12	&0.08	&2	&45	\\
        J0451	&0.28	&0.05	&2700	&1500	&0.20	&0.10	&6	&46	\\
        J0647	&0.28	&0.01	&3500	&400	&0.07	&0.02	&11	&36	\\
	J0717	&0.26	&0.07	&5700	&4000	&0.07	&0.04	&-58	&29	\\
	J0744	&0.74	&0.03	&2000	&300	&0.59	&0.02	&-0.6	&1	\\
	J0911	&0.39	&0.19	&850	&780	&0.29	&0.10	&-17	&34	\\
	J1149	&0.23	&0.02	&3900	&1100	&0.10	&0.03	&-65	&17	\\
	J1423	&0.29	&0.07	&1700	&1000	&0.34	&0.09	&-25	&8	\\
	J2129	&0.31	&0.06	&3700	&2600	&0.07	&0.04	&-2	&40	\\
	J2214	&0.29	&0.07	&2000	&1200	&0.32	&0.10	&54	&20	\\
      \hline
\end{tabular}
\label{tab_NFW_2D_par}
\end{table}

\begin{figure}
       \resizebox{\hsize}{!}{\includegraphics{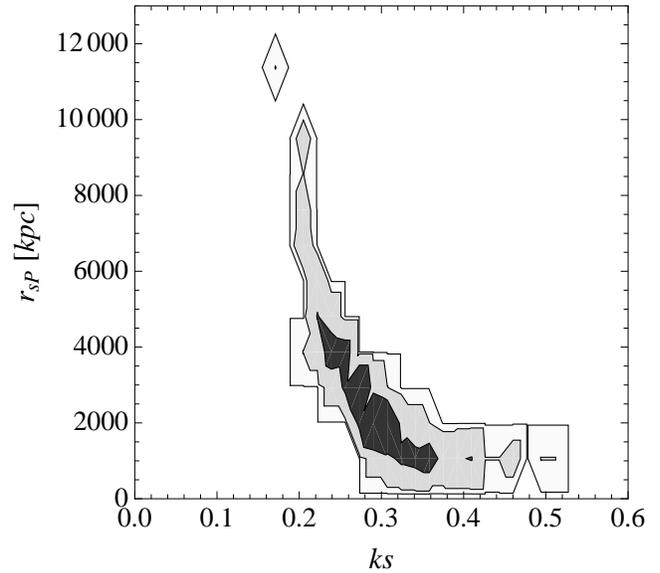}}
       \caption{Contour plot of the marginalised PDF for the strength $\kappa_\mathrm{s}$ and the projected radius $r_\mathrm{sP}$ for MACSJ0018. Contours are plotted at fraction values $\exp (-2.3/2)$, $\exp(-6.17/2)$, and $\exp(-11.8/2)$ of the maximum, which denote confidence limit region of 1, 2 and $3\sigma$ in a maximum likelihood framework, respectively.}
	\label{fig_MACSJ0018_PDF_ks_rsP}
\end{figure}

We performed strong lensing analyses of the inner regions of the high redshift MACS sample. A detailed description of the strong lensing features of the sample can be found in \citet{zit+al11}. We adopted a two step procedure. Firstly, we obtained the surface mass density for each cluster. Secondly, we modelled the map with a projected NFW profile.

We obtained a pixelated map of the surface mass density of each cluster by use of the PixeLens software \citep{sa+wi04}. PixeLens generates a statistical ensemble of models which can exactly reproduce the positions and parities of all multiple-image. The ensemble automatically explores degeneracies in convergence and provides uncertainties. For each cluster, we computed 200 convergence maps. Parameters set to run the code are listed in Table~\ref{tab_PixeLens}. For each lens, we analysed a circular area centred on the brightest cluster galaxy (BCG). The external radii are listed in Table~\ref{tab_PixeLens}; they extend just behind the outermost image. These criteria select a region larger than that enclosed by the Einstein radius, see Tables~\ref{table:sample} and~\ref{tab_PixeLens}. The effectiveness of PixeLens is consequently poorer in the outer pixels. We then excised these pixels as explained in the following. In determining the optimal pixel resolution we had to counterbalance two competing effects. On one side, a more refined grid of pixels allows for a better constraining of small scale structures. On the other side, a too large number of pixels slows the code to a not sustainable extent. It can also induce over-sampling effects. From the analysis of lensing in simulated systems, we found that a radius of $\sim 14-15$ pixels gives the best compromise. The explicit resolutions are listed in Table~\ref{tab_PixeLens}.

PixeLens may offer a poor description of the mass profile in the very outer pixels, with an unphysical sharp drop in density just outside the Einstein radius. We preliminary computed the slope profile of the surface density in spherical annuli centred in the denser pixel and checked for logarithmic density slope values smaller than -2, the minimum allowed slope for a projected NFW profile. We required at least two consecutive annuli with slope lower than the threshold. We fixed the minimum convergence at the averaged convergence of the first spherical region with unphysical slope and excluded from the analysis pixels with convergence lower than the minimum. With the aid of simulated images, we checked that this cutting procedure fares very well. In the radial span of just a couple of pixels near the Einstein radius, the logarithmic slope suddenly falls from $\gs -1$ in the internal region (where $\xi \ll r_\mathrm{sP}$) to values $\ls -3$ or even steeper. We retrieved this feature, which is artefact of PixeLens, in both simulated or observed clusters. We checked that the value of the minimum allowed convergence was negligibly affected by the use of different cutting criteria, such as either a different minimum value for the slope ($\gs -1.5$) or by requiring that the allowed slope variation between adjacent annuli was smaller than a maximum difference.

We also excluded the central pixels. \citet{ume+al11a,ume+al11b} found a small offset of typically $\ls d_\mathrm{off}=20~\mathrm{kpc~h}^{-1}$ between the BCG and the dark matter centre of mass recovered from strong lens modeling. Even if in PixeLens the centre position is left free and only a preliminary geometrical centre (usually coinciding with the position of the BCG) is used, we conservatively limited our analysis to radii greater than 2 $d_\mathrm{off}$ from the pixel with the maximum convergence, beyond which the cluster miscentering effects are negligible. For $z>0.5$, this criterium usually cuts only the very central denser pixel. The exclusion of the central pixels is also conservative with regard to the effects of baryons on the total matter profile. In fact, in the investigated range the total distribution is dominated by dark matter, whose behavior is very well modelled in numerical simulations. An example of pixels which passed the cuts is plotted in Fig.~\ref{fig_MACSJ0018_density_map}.

As a second step, we modelled the pixelated maps with a projected NFW profile. We looked for the minimum of
\begin{equation}
\label{eq_fit_1}
\chi^2 = \sum_i \left[ \kappa_i -\kappa_\mathrm{NFW}(x_1^i,x_2^i;\theta_{0,1}, \theta_{0,2},\kappa_\mathrm{s}, r_\mathrm{sP},\epsilon,\theta_\epsilon)\right]^2
\end{equation}
where the sum runs over the pixels and $k_i$ is the measured convergence of the $i$-th pixel of coordinates $x_1^i,x_2^i$; $\kappa_\mathrm{NFW}$ is the theoretical prediction. Each pixel was given the same weight. We fitted 6 free parameters: the halo centre coordinates, $\theta_{0,1}$ and $\theta_{0,2}$; the strength $\kappa_\mathrm{s}(>0)$ and the projected radius $r_\mathrm{sP}(>0)$; the ellipticity $(0<)\epsilon(\leq1)$ and the orientation angle $\theta_\epsilon$. In Fig.~\ref{fig_MACSJ0018_density_map} we plot the result of such fitting procedure for a mean convergence map.

We carried out the fitting procedure for each map. From the derived ensemble of maximum likelihood parameters, we obtained the posteriori distribution of projected NFW parameters. For each map, we considered only the best-fit parameters. Since we fitted each map separately, correlation effects and convergence stability for each pixel were not considered at the level of Eq.~(\ref{eq_fit_1}). These effects were taken into account later by grouping the sets of maximum likelihood parameters. An example of fitting results is plotted in Fig.~\ref{fig_MACSJ0018_density_map}.

Our approach is alternative to standard ones which constrain NFW elliptical profiles by directly fitting multiple image systems with parametric models. Due to the presence of local substructures, such a method might underestimate parameter errors. On the other hand, we fitted the 2D mass distribution rather than the image positions. Lensing is highly non-linear. Attempting to directly fit the multiple images positions with an NFW model can then give different results from those obtained by reproducing the image positions with a pixelated map and then fitting the map with the same parametric model. This is more similar to NFW fitting in numerical simulations, where the 3D mass density is directly fitted.

Fitting results for the relevant parameters are summarised in Table~\ref{tab_NFW_2D_par}. The very large values for $r_\mathrm{sP}$ and the related very large errors reflect the inability of the fitting procedure to effectively constrain the scale length. Strong lensing focuses on the very inner regions whereas a precise determination of $r_\mathrm{sP}$  would require a coverage up to larger radii, as in weak lensing analyses. Degeneracies effects make even very large values of $r_\mathrm{sP}$ compatible with multiple image positions.  As a consequence, the central momentum of the distribution is pushed to large values too. This can be seen in the marginalized PDF distribution for $\kappa_\mathrm{s}$ and $r_\mathrm{sP}$, see Fig.~\ref{fig_MACSJ0018_PDF_ks_rsP}. Most likely values for $r_\mathrm{sP}$ are at small values but the tail at large values is really extended. The large uncertainty on $r_\mathrm{sP}$ will carry over into a large uncertainty on halo mass and concentration.

\section{Deprojection}
\label{sec_depr}

\begin{table}
\centering
\caption{NFW intrinsic parameters of the $z>0.5$ MACS sample as obtained from the Bayesian deprojection. Columns 2, 3 and 4 list the values of mass, concentration and (cosine of) inclination angle, respectively. Central values and dispersions are computed as bi-weight estimators of the marginalised PDFs.}
\begin{tabular}[c]{cr@{$\,\pm\,$}lr@{$\,\pm\,$}lr@{$\,\pm\,$}l}
        \hline
        \noalign{\smallskip}
        MACS	&	\multicolumn{2}{c}{$M_{200}$} &	\multicolumn{2}{c}{$c_{200}$} & \multicolumn{2}{c}{$\cos \theta$}      \\
         		&	\multicolumn{2}{c}{$[10^{15}M_\odot/h]$} &	\multicolumn{2}{c}{} & \multicolumn{2}{c}{}      \\
	\hline
	J0018 &4.1 &2.7 &1.9 &0.9 &0.59 &0.29 \\
	J0025 &4.3 &3.2 &1.6 &1.1 &0.73 &0.27 \\
	J0257 &2.7 &1.7 &2.7 &1.1 &0.76 &0.26 \\
	J0451 &4.8 &2.5 &1.6 &0.7 &0.74 &0.25 \\
	J0647 &6.8 &1.4 &1.0 &0.2 &0.81 &0.12 \\
	J0717 &4.5 &3.0 &1.5 &1.2 &0.57 &0.20 \\
	J0744 &8.0 &1.3 &3.1 &0.5 &0.38 &0.23 \\
	J0911 &1.8 &1.7 &3.6 &2.2 &0.57 &0.30 \\
	J1149 &5.1 &1.9 &0.9 &0.3 &0.85 &0.11 \\
	J1423 &2.9 &2.2 &2.0 &0.9 &0.59 &0.29 \\
	J2129 &3.5 &3.1 &1.7 &1.1 &0.80 &0.12 \\
	J2214 &3.2 &2.8 &2.3 &1.2 &0.61 &0.29 \\
\hline
\end{tabular}
\label{tab_NFW_3D_par}
\end{table}

Deprojecting a surface density map to infer the intrinsic 3D shape is an under-constrained astronomical problem. Apart from 3 coordinates fixing the halo centre, the 3D NFW halo features seven parameters: the density profile is described by two parameters ($M_{200}$ and $c_{200}$), the shape by two axial ratios ($q_1$ and $q_2$), the orientation by three angles, $\theta$ and $\varphi$ and $\psi$. On the other hand, since lensing is dependent on the projected mass density, from lensing observations we only get four constraints relating the intrinsic parameters: the lensing strength $\kappa_\mathrm{s}$, the projected radius $r_\mathrm{sP}$, the ellipticity $\epsilon$ and the orientation $\theta_\epsilon$.
Even combining multi-wavelength observations, from X-ray through the optical and to radio bands, one can only constrain the elongation of the cluster along the line of sight \citep{ser07}.

To assess realistic probability distributions for the intrinsic parameters we performed a statistical Bayesian analysis. The use of some a priori hypotheses on the cluster shape can help to disentangle degeneracies. We applied some methods already employed in gravitational lensing analyses \citep{ogu+al05,cor+al09,ser+al10}. In particular, we followed \citet{se+um11}, where further details on the method can be found. The Bayes theorem states that
\beq
\label{baye}
p(\bfP | \bfd) \propto {\cal L}( \bfP|\bfd) p(\bfP),
\eeq
where $p(\bfP | \bfd)$ is the posterior probability of the parameters $\bfP$ given the data $\bfd$, ${\cal L}( \bfP|\bfd)$ is the likelihood of the data given the model parameters, and $p(\bfP)$ is the prior probability distribution for the model parameters.

Using the functional dependence of the projected parameters on the intrinsic ones, the likelihood for the intrinsic parameters is the likelihood ${\cal L}(\kappa_\mathrm{s}, r_\mathrm{sP}, \epsilon, \theta_\epsilon)$ derived in Sec.~\ref{sec_stron}. This distribution was smoothed using a Gaussian kernel estimator \citep{vio+al94,ryd96}.

As prior for the axial ratios $q_1$ and $q_2$, we considered the $N$-body predictions in Eqs.~(\ref{nbod3}--\ref{nbod4}). We always put a lower bound $q_1 \ge 0.1$. For the alignment angle $\theta$ and the the azimuth angle $\varphi$, we considered a random distribution. For the mass, we always used a flat prior $p(M_{200}) = const.$ in the range $10^{13}\ \le M_{200}/ (M_\odot h^{-1}) \le 10^{16}$, whereas the prior for the concentration was flat in the range $0 < c_{200} \le 30$ and null otherwise. Posterior PDFs were computed by running four Markov chains. We checked for chain convergence by verifying that the standard var(chain mean)/mean(chain var) indicator was less than 1.1. We computed at least 10000 samples per chain and eventually added ten thousands more until the convergence criterium was satisfied.

Results are summarised in Table~\ref{tab_NFW_3D_par}. Central location and dispersion for each parameter are computed as biweight estimators of the corresponding marginalised posterior probability function (PDF) \citep{bee+al90}. Interesting results can be obtained on the mass, concentration and elongation of the clusters. Since we exploited only strong lensing data in the inner regions, we cannot determine the virial halo mass accurately. Our typical error on $M_{200}$ is then $\gs 60$ per cent. This reflects the large uncertainty in the projected scale radius, discussed in Sec.~\ref{sec_stron}. A better determination would require either weak lensing data at larger radii or a much larger number of multiple image systems. The statistical uncertainty in the concentration measurement is large too ($\ls 50$ per cent), but better than the mass accuracy.

Dealing with a triaxial analysis, we have five more parameters (three orientation angles and two axial ratios) with respect to a spherical analysis. The errors on $M_{200}$ and $c_{200}$ are consequently larger than when assuming spherical symmetry \citep{se+um11}. Elongation can be somewhat constrained since it is directly connected to the lensing strength. On the other hand, the PDFs of the intrinsic axial ratio $q_1$ and $q_2$ and of the angle $\varphi$ are dominated by the a priori assumptions. Our strong lensing data are not constraining in this regard and we do not discuss them. The posterior distribution of $\psi$ just reflects the likelihood on $\theta_\epsilon$ and is not significant in this context.

The final balance is that, with respect to a standard spherical analysis, we can constrain a third parameter, i.e., the elongation, at the expense of a poorer accuracy on halo mass and concentration. Our Bayesian analysis also cuts unphysical models, such as those corresponding to extremely large values of $r_\mathrm{sP}$ which are still represented by the likelihood.

\section{Results}\label{sec_results}

\subsection{Mass and concentration}

\begin{figure}
       \resizebox{\hsize}{!}{\includegraphics{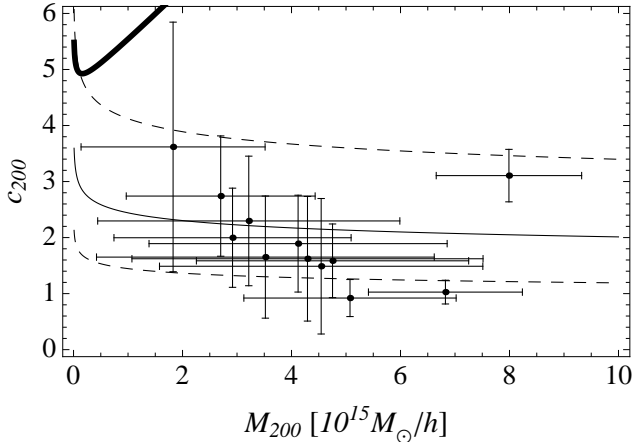}}
       \caption{Estimated mass and concentration of the $z>0.5$ MACS clusters versus theoretical predictions. The full and dashed thin lines denote the central value and the 1$\sigma$ predictions, respectively, for the full sample of clusters in \citet{duf+al08}. The redshift of the simulated cluster population is fixed at the average redshift of the observed sample, $z \simeq 0.56$; the thick line is the prediction from \citet{pra+al11}.}
	\label{fig_M200_c200}
\end{figure}

\begin{figure}
       \resizebox{\hsize}{!}{\includegraphics{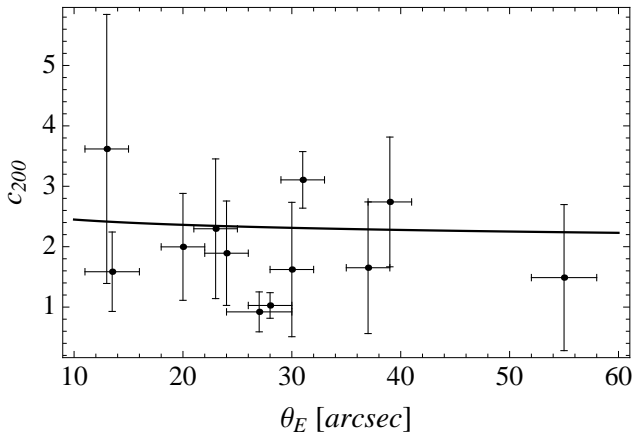}}
       \caption{Effective Einstein radii versus concentrations for the $z>0.5$ MACS clusters.The full line is the theoretical prediction for unrelaxed clusters at $z\simeq 0.56$. Concentrations for relaxed clusters are expected to be significantly higher.}
	\label{fig_theta_E_c200}
\end{figure}

The estimated mass-concentration relation for the high redshift MACS sample is plotted in Fig.~\ref{fig_M200_c200}. Our strong lensing analysis could naturally not determine the total cluster masses very accurately, whereas the constraints on the concentrations were slightly tighter. However, the over-concentration problem can be addressed even if uncertainties are large. We found out that the $z>0.5 $ MACS clusters are very massive as expected ($M_{200} \gs 10^{15}M_\odot/h$), with quite low concentrations $c_{200}\ls 3$. Usually, strong lensing selected clusters with mass of this order were found to be over-concentrated to the extent of $c_{200}\sim 10$. Even with a quite large error of $\delta c_{200}\ls 1.2$ we can then distinguish clusters in agreement with the theoretical $c(M)$ relation from outliers. Furthermore,  most of the theoretical studies on the $c(M)$ relation predict a quite shallow dependence of the concentration on the mass  \citep{net+al07,mac+al08,gao+al08,duf+al08}, so that comparison with our results can be meaningful even with large uncertainties $\delta M_{200}$. However, large errors prevent us from determining the slope in the observed $c(M)$ relation.

We find a noteworthy agreement between observations and predictions for the MACS sample. All of the measured concentrations are compatible within 1$\sigma$ with the predicted values. A small decrement of concentrations with increasing masses seems to be in order, but large errors asks for caution.

Recently, \citet{pra+al11} noticed a flattening and upturn of the relation with increasing mass and redshift. Predicted concentrations for the MACS sample would be significantly higher, see Fig.~\ref{fig_M200_c200}. In that case, our results would pose a under-concentration problem.

The concentration of relaxed, lower-$z$ clusters, is expected to correlate with the Einstein radius \citep{sa+re08}. In fact, more concentrated clusters should have more mass in the middle, thus boosting their Einstein radius. We did not recover this trend in our higher redshift sample, see Fig.~\ref{fig_theta_E_c200}. The overplotted theoretical $c(\theta_\mathrm{E})$ relation has been obtained from the $c(M)$ derived in \citet{duf+al08} by exploiting the expected trend between masses and Einstein radii found for unrelaxed clusters \citep[$\theta_\mathrm{E} \propto M_{200}^{1.6}$,][]{br+ba08}. Strong lensing analyses can measure very accurately the Einstein radius, whereas the determination of the concentration is less precise. Large errors in $c_{200}$ might then hide the real trend in the $c(\theta_\mathrm{E})$ relation for the $z>0.5$ MACS sample but apparently here the Einstein radius is not larger for smaller concentrations, as expected for unrelaxed high redshift massive halos.

\subsection{Orientation}

\begin{figure}
       \resizebox{\hsize}{!}{\includegraphics{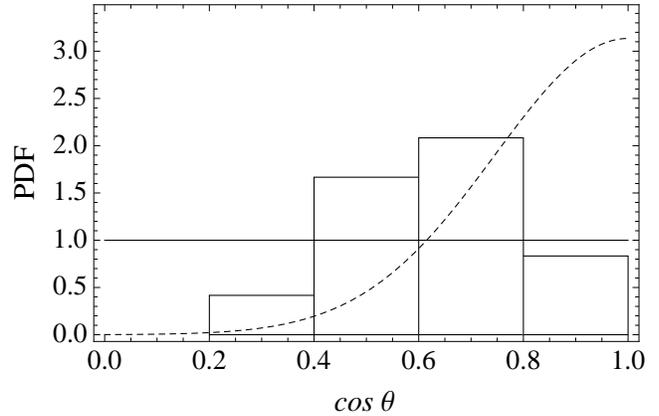}}
       \caption{Theoretical probability density functions for the orientation ($\cos \theta$) of randomly oriented (full line) or biased ($\sigma_{\cos \theta}=0.115$, dashed line) $N$-body like clusters versus measured values for the $z>0.5$ MACS sample (normalized histogram). The predicted distributions were convolved with a Gaussian distribution with dispersion of $\sim 0.23$.}
	\label{fig_PDF_cos_theta}
\end{figure}

\begin{figure}
       \resizebox{\hsize}{!}{\includegraphics{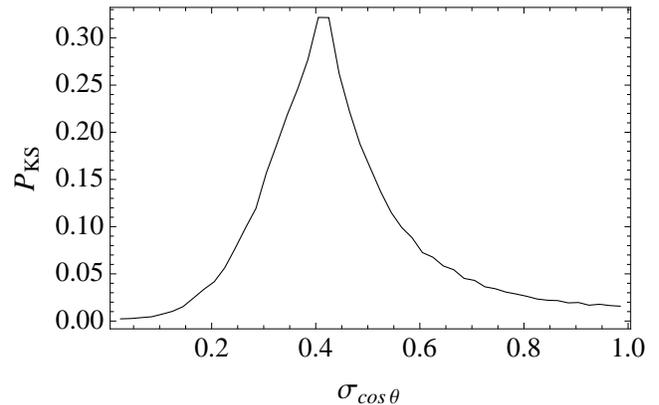}}
       \caption{Kolmogorov-Smirnov significance level that the orientations of the $z>0.5$ MACS sample are drawn from a biased population of clusters with dispersion $\sigma_{\cos \theta}$. The predicted distributions were convolved with a Gaussian function with dispersion of $\sim 0.23$.}
	\label{fig_KS_sigma_cos_theta}
\end{figure}

Very low values of the orientation angles $\theta$ for the MACS sample would suggest an orientation bias. Measured values are listed in Table~\ref{tab_NFW_3D_par}. Predicted versus observed values are plotted in Fig.~\ref{fig_PDF_cos_theta}. To account for the observational uncertainty, theoretical predictions were convolved with a Gaussian function with dispersion equal to the average uncertainty on $\cos \theta$ ($\sim 0.23$). With respect to a randomly oriented sample, there is an excess at low orientation angles, $\cos \theta \gs 0.5$. The more inclined clusters ($\cos \theta \gs 0.8$) are those which appear nearly circular in the plane of the sky ($\epsilon \ls 0.1$).

The Kolmogorov-Smirnov (KS) test can provide further insight. Neither of the two extreme cases considered (strong lensing orientation bias with $\sigma_{\cos \theta}=0.115$ or random orientations)  provide a very good description of the data but the strongly biased case is slightly preferred. The KS significance level is of 0.9 per cent for the biased case or 0.4 per cent for the random case.  To assess the size of any orientation bias, we run the KS test for different values of $\sigma_{\cos \theta}$, see Fig.~\ref{fig_KS_sigma_cos_theta}. The significance level is maximum for $\sigma_{\cos \theta}=0.4$, suggesting that a mild form of orientation bias affects the X-ray selected MACS sample.

\subsection{Elongation and ellipticity}

\begin{table}
\centering
\caption{Inferred elongations $e_\Delta$ of the $z>0.5$ MACS sample. The lower $e_\Delta$, the more prominent the elongation along the line of sight. Central values and dispersions are computed as bi-weight estimators of the marginalised PDFs.}
\label{tab_e_Delta}
\begin{tabular}[c]{cr@{$\,\pm\,$}l}
        \hline
        \noalign{\smallskip}
        MACS	&	\multicolumn{2}{c}{$e_\Delta$}    \\
	\hline
	J0018 & 1.6 & 0.6\\
	J0025 & 1.3 & 0.6\\
	J0257 & 1.2 & 0.5\\
	J0451 & 1.3 & 0.5\\
	J0647 & 1.0 & 0.3\\
	J0717 & 0.9 & 0.4\\
	J0744 & 2.1 & 0.6\\
	J0911 & 1.5 & 0.5\\
	J1149 & 0.9 & 0.4\\
	J1423 & 1.6 & 0.6\\
	J2129 & 0.9 & 0.4\\
	J2214 & 1.5 & 0.6\\
\hline
\end{tabular}
\end{table}

\begin{figure}
       \resizebox{\hsize}{!}{\includegraphics{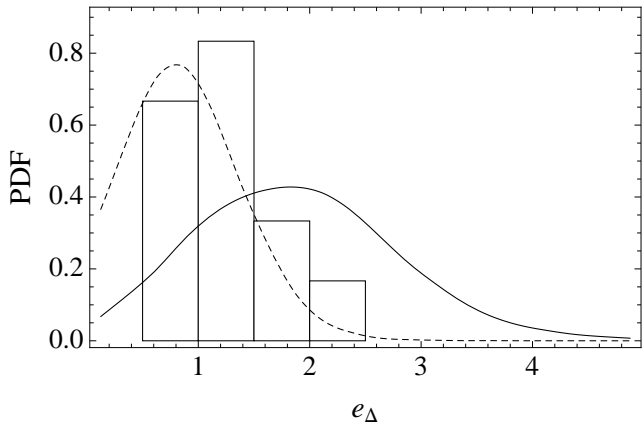}}
       \caption{Theoretical probability density functions for the elongations of either randomly oriented (full line) or biased ($\sigma_{\cos \theta}=0.115$, dashed line) $N$-body like clusters versus measured values for the $z>0.5$ MACS sample (normalized histogram).}
	\label{fig_PDF_e_Delta}
\end{figure}

\begin{figure}
       \resizebox{\hsize}{!}{\includegraphics{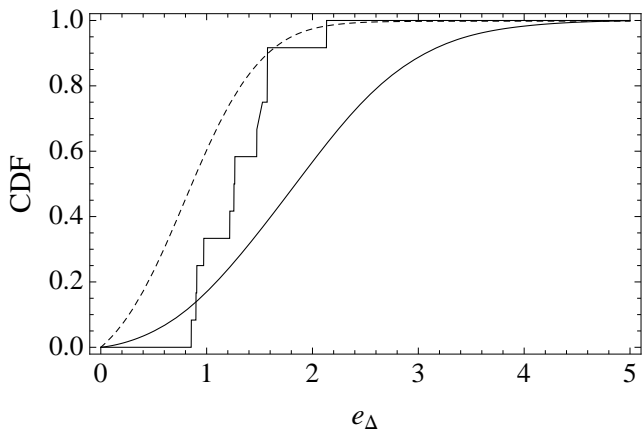}}
       \caption{Predicted cumulative distribution functions for elongation versus measurements of the $z>0.5$ MACS sample. The step-line is for the observed sample. The smooth full and dashed lines are the predicted distributions for  $N$-body like clusters with either random or biased ($\sigma_{\cos \theta}=0.115$) orientations, respectively.}
	\label{fig_CDF_e_Delta}
\end{figure}

\begin{figure}
       \resizebox{\hsize}{!}{\includegraphics{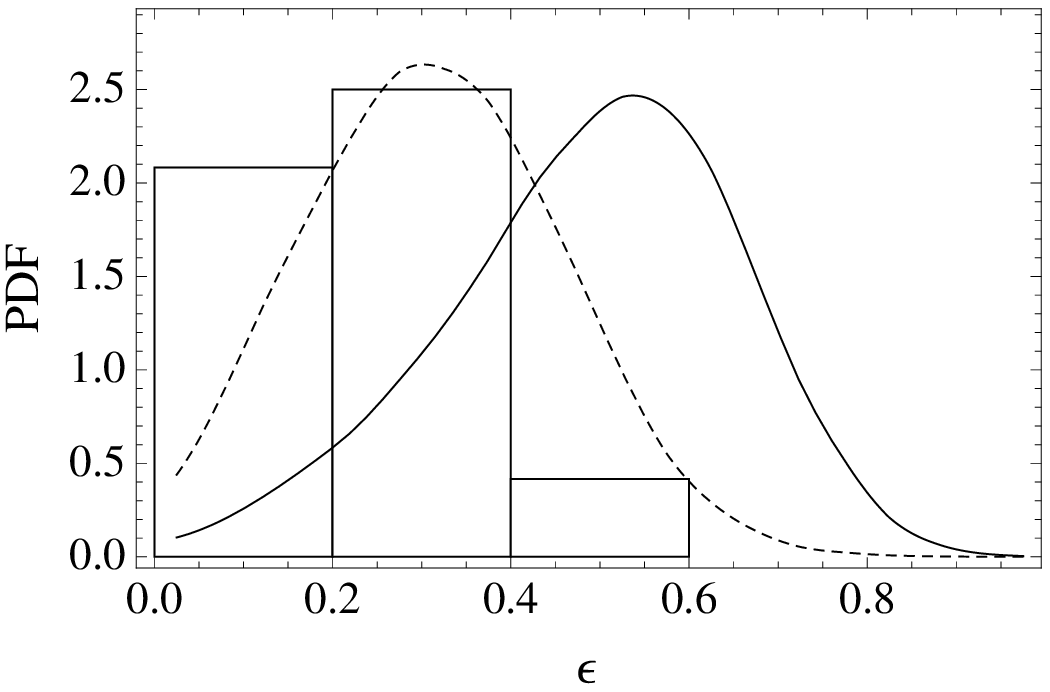}}
       \caption{The same as Fig.~\ref{fig_PDF_e_Delta} for the projected ellipticity.}
	\label{fig_PDF_epsilon}
\end{figure}

\begin{figure}
       \resizebox{\hsize}{!}{\includegraphics{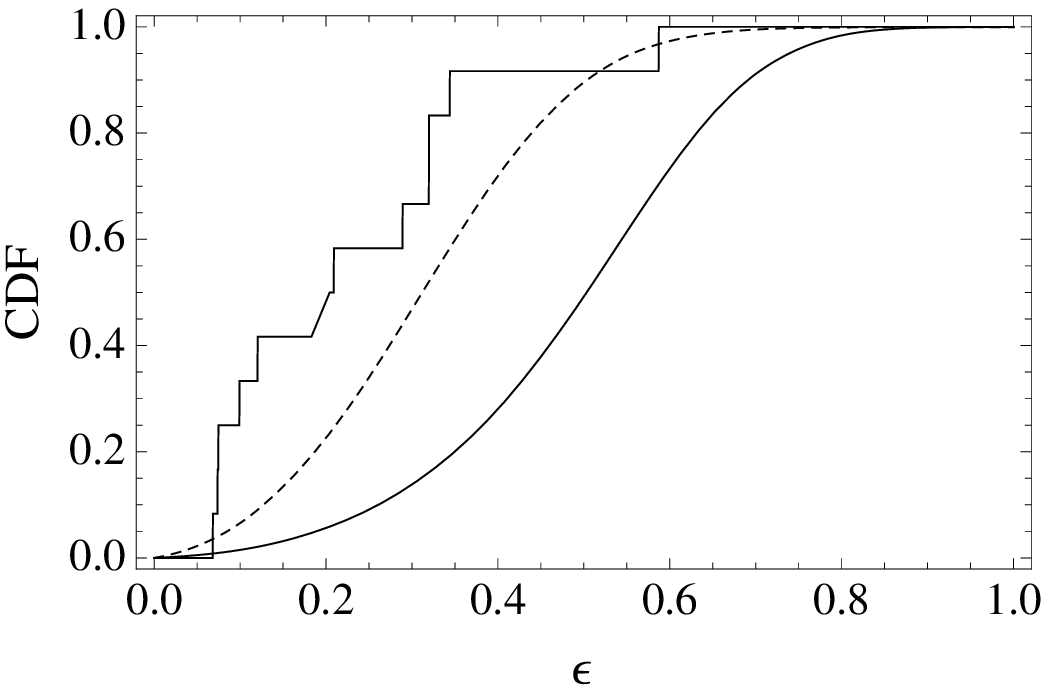}}
       \caption{The same as Fig.~\ref{fig_CDF_e_Delta} for the projected ellipticity.}
	\label{fig_CDF_epsilon}
\end{figure}

\begin{figure}
       \resizebox{\hsize}{!}{\includegraphics{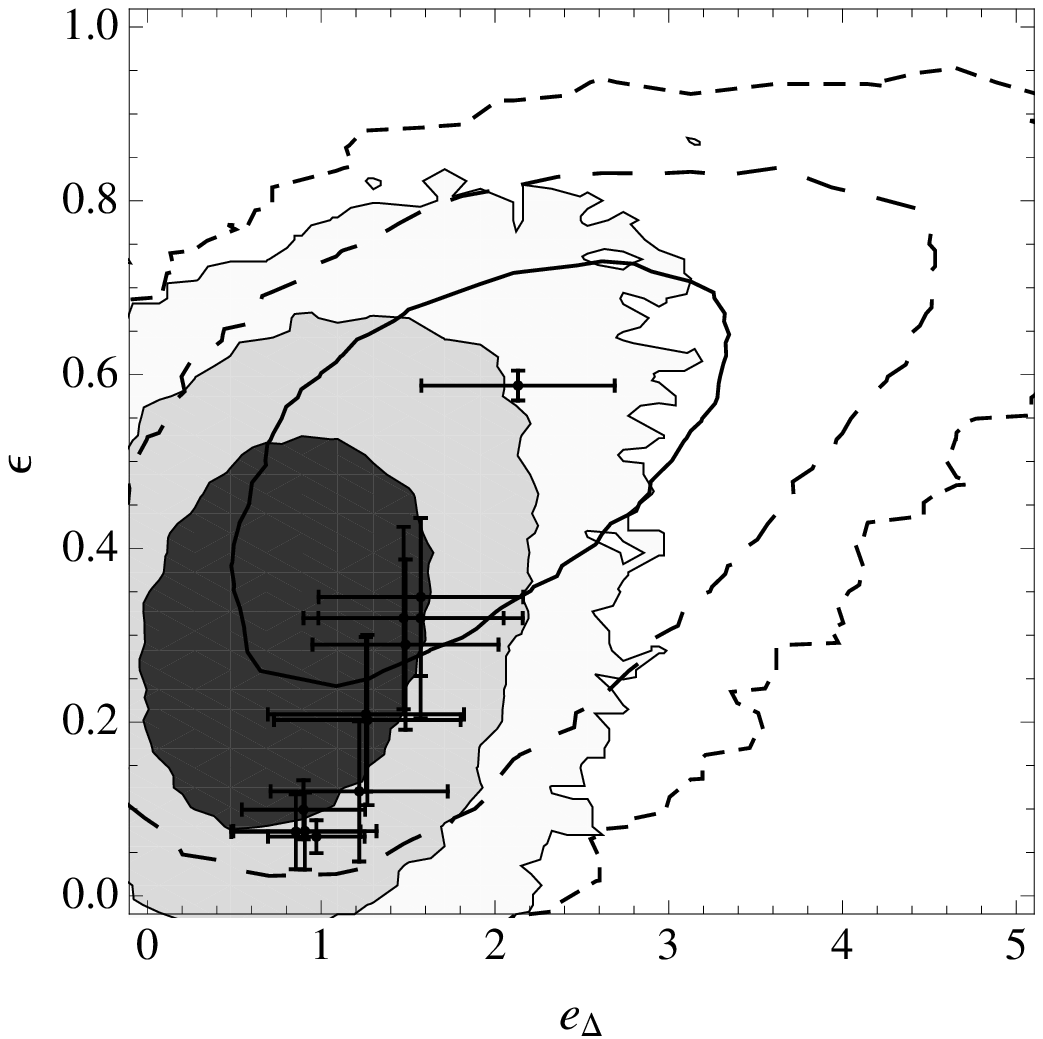}}
       \caption{Ellipticities and elongations of the MACS sample (crosses) versus the theoretical predictions. Contours are plotted at fraction values $\exp (-2.3/2)$, $\exp(-6.17/2)$, and $\exp(-11.8/2)$ of the maximum, which denote confidence limit region of 1, 2 and $3\sigma$ in a maximum likelihood framework, respectively. The shadowed regions are for a population of $N$-body like haloes with a biased orientation ($\sigma_{\cos \theta}=0.115$); the thick contours are for randomly oriented clusters.}
	\label{fig_e_Delta_epsilon}
\end{figure}

A combined analysis of elongations and ellipticities can provide further information on the orientation bias. Estimated elongations are listed in Table~\ref{tab_e_Delta}. The observed versus the predicted values of the elongations are plotted in Fig.~\ref{fig_PDF_e_Delta}. The cumulative distributions are plotted in Fig.~\ref{fig_CDF_e_Delta}. As theoretical distributions we considered a population of clusters with $N$-body like axial ratios. The masses and the redshifts of the theoretical distribution are fixed to the average values of the $z>0.5$ MACS sample. As before, we considered two cases for the orientation: either randomly oriented lenses or clusters strongly inclined toward the line of sight ($\sigma_{\cos \theta}=0.115$). Theoretical distributions were smoothed to account for observational uncertainties by convolving with a Gaussian function with dispersion equal to the average observational error on the measured elongations, i.e. $\sim 0.5$. The biased population makes a better job in reproducing observed measured elongations, with a KS significance level of 0.3 per cent compared to 0.1 per cent for the randomly oriented clusters.

\citet{men+al11} found that three clusters in the sample, namely MACSJ0717, MACSJ0025, and MACSJ2129, have extremely large lensing cross sections. Triaxiality might help to understand these outliers. In fact, two out the the three clusters, i.e. MACSJ0717 and MACSJ2129, have two of the smaller $e_\Delta$ values ($\sim 0.9$) in the sample. Convergence is proportional to $1/e_\Delta$, and elongation can enhance the cross section. MACSJ0717 is known to sit at the tip of a prominent filament, which further supports the view of an elongated structure.

The same analysis can be applied to the measured ellipticities, see Table~\ref{tab_NFW_2D_par}. Note that $\epsilon$ is directly inferred from the strong lensing analysis whereas $e_\Delta$ was estimated through the deprojection method. The observed versus the predicted values of the ellipticities are plotted in Fig.~\ref{fig_PDF_epsilon}. The cumulative distributions are plotted in Fig.~\ref{fig_CDF_epsilon}. We considered the same theoretical distributions as for the elongations. This time, the convolving normal function accounting for observational uncertainties had a dispersion of $\sim 0.07$. You can see by eye that the biased population can reproduce the observed values much better than the random one, in particular it predicts the observed excess at small values. In more quantitative terms, the KS significance level for the biased population is a robust 10 per cent, whereas for the random sample we got a negligible $2\times10^{-4}$ per cent.

A further element in favour of an orientation bias is given by the analysis of the two-dimensional distributions, see Fig.~\ref{fig_e_Delta_epsilon}. Randomly oriented or biased populations in our analysis share the same axial ratio distribution. On the other hand, the orientation bias is compatible with a significant number of inclined clusters which are at the same time nearly circular when projected in the plane of the sky (small values of $\epsilon$) and quite elongated along the line of sight (small values of $e_\Delta$). This feature is seen in the distribution of observed values. All pairs of measured $\epsilon$ and $e_\Delta$ are compatible within the 2$\sigma$ confidence level with the biased population.

\section{Systematics}
\label{sec_syst}

\begin{figure}
       \resizebox{\hsize}{!}{\includegraphics{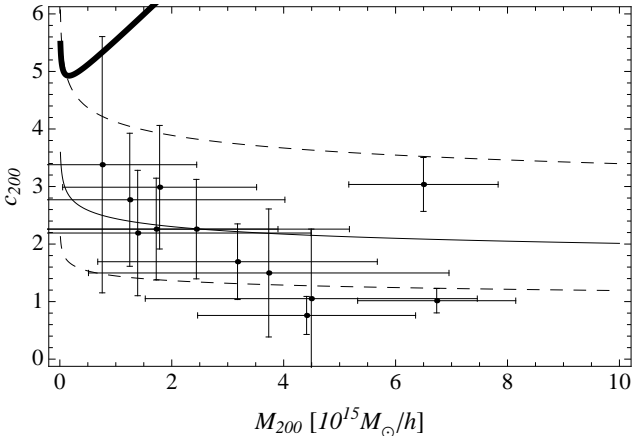}}
       \caption{Estimated mass and concentration of the $z>0.5$ MACS clusters versus theoretical predictions assuming priors for mass and concentration flat in logarithmic bins. Symbols and styles are the same of Fig.~\ref{fig_M200_c200}.}
	\label{fig_M200_c200_logprior}
\end{figure}

Let us discuss some systematic uncertainties that might affect our analysis. The well-established modelling method used in \citet{zit+al11} to identify many sets of multiply-lensed images has been very successful in determining source redshifts. Furthermore, within the PixeLens approach, statistical uncertainties in the modelling due to unknown source redshifts (as far as the error on $z_\mathrm{s}$ is $\ls 0.5$) are lower than degeneracies in the parameter space due to variations in the mass density \citep{sa+wi04}. Effects of redshift uncertainties are then expected to be negligible.  Firstly, over-(under) estimating the source redshift would decrease the lensing strength of the lens, and in turn the estimations of mass and concentration. This trend would not solve any over-concentration problem in $\Lambda$CDM, which would be artificially solved only for mass and concentration systematically changed in opposite directions. Secondly, the main effect of redshift uncertainties would be on the determination of the profile slope. Here, we were interested in comparison with numerical simulations which assumed a NFW profile for the matter halo. Then, we could keep the profile slope fixed and we did not determine it from the data. Thirdly, any variation of $\delta z_\mathrm{s} \sim 0.5$ for a typical lensed source at $z_\mathrm{s} \sim 2$ would change the image position by 6-10 per cent. At a typical Einstein radius of $30\arcsec$, this would translate into an astrometric error of $\ls 2$-$3\arcsec$. Since we used a pixelated map to reproduce the mass density, the effect on parameter degeneracy is then negligible. In fact, our typical pixel resolution for the $z>0.5$ MACS clusters is of $\sim 3-4\arcsec$, see Table~\ref{tab_PixeLens}, which is worse than the redshift uncertainty.

Substructure and line-of-sight haloes might also play a role \citep{da+na11}. A group-sized halo residing slightly behind the lens typically brings about an astrometric shift of $\sim 0.25\arcsec$, with largest deviations of $\sim 1\arcsec$. If the masses of the two line-of-sight structures are roughly equivalent the deviations increase to $\sim0.9 \arcsec$ on average and can be as high as a few arcseconds. The effect is more dramatic for uncorrelated line-of-sight structures which should perturb the image locations by $\ls 1 \arcsec$ for a lens at $z_\mathrm{s} \sim 0.5$. These effects turns then out to be negligible when compared with our typical pixel resolution.

A different source of error might come from the use of priors in the Bayesian analysis. If the likelihood is not peaked, final result might reflect the employed a priori hypotheses. Whenever the parameter values are limited
from above and below, as it is conceivably for the mass of MACS clusters, which are known to be very massive, and the concentration, which by definition is positive and very hardly exceeds very large values, flat priors in the allowed range, as the ones we used in Sec.~\ref{sec_depr}, are the best choice. However, an alternative might be to use priors uniform in logarithmically-spaced decades, as usually done for parameters with only lower bounds. We then re-run the analysis with these alternative priors for mass and concentration. Results are summarised in Fig.~\ref{fig_M200_c200_logprior}. The agreement with the results obtained in Sec.~\ref{sec_depr} shows that our results were determined by the data and that the role of priors was secondary. The only effect of logarithmically-spaced uniform priors was to favor slightly smaller masses. The shift was however much smaller than the estimated uncertainty, so that the determined trend for the $c(M)$ relation is unaffected.

\section{Conclusions}
\label{sec_conc}

The full understanding of the interplay between mass and concentration in galaxy clusters is an open problem. The rich strong lensing features of the $z>0.5$ MACS clusters offer the opportunity to study the $c(M)$ relation in a well defined statistical sample. We exploited a full triaxial lensing analysis, where the matter halos were modelled as triaxial NFW ellipsoids with arbitrary orientations. The method is not affected by systematics related to shape, so that the resulting estimated concentrations should not suffer from any orientation bias.

Our results are apt to comparison with $N$-body simulations. We find that the $c(M)$ relation of the $z>0.5$ MACS sample is in remarkable agreement with theoretical predictions. There is no hint to an over-concentration problem. The clusters turn out to be preferentially elongated along the line of sight, although this orientation bias is much smaller than for strong-lensing selected clusters.

The obtained masses and concentrations are also in line with some general expectations based on the properties of the sample. The MACS clusters are X-ray selected. Gas distribution is usually rounder than the matter density, so triaxiality plays a smaller role in cluster selections \citep{gav05}. The orientation bias as well as the over-concentration problem should be then sensibly reduced with respect to strong-lensing selected samples. On the other hand, the $z>0.5$ MACS clusters are representative of the very high mass tail of the full cluster distribution, so that the role of a smaller bias might be amplified. We indeed retrieve such expected trends.

With more and more reliable observational estimates of the $c(M)$ relation, the onus seems now to be on the theoretical predictions from $N$-body analyses. Dark matter $N$-body simulations usually samples very few massive clusters and their predictions at the high-redshift and massive end are mainly based on extrapolations of results of low $z$ haloes. Results from independent analyses are still in disagreement and estimated concentrations for massive clusters can differ by $\ls 50$ per cent \citep{pra+al11}. Baryonic physics can play an important role too, especially in the very inner regions probed by strong lensing. The baryonic contribution to the assembling of halos in large simulation volumes has yet to be fully understood. The explicit effects of cooling, feedback, and baryonic-DM interplay on the mass profile and concentration are also still ambiguous \citep[e.g.,][]{gne+al04,Roz+al08,Duffy10}. Continuous advancements in computational power may soon enable sufficiently high-resolution description to enlighten the true (theoretical) $c(M)$ relation for the most massive halos.

Although we have found no discrepancy between theory and observations in the $c(M)$ relation of the high-$z$ MACS sample, we note that this sample appears to consist of mostly unrelaxed clusters. The high X-ray luminosity, along with the relatively high redshift and recent SZ effect images (via private communication), are in support of this claim, in addition to noticeable substructure seen in optical and lensing analyses \citep{zit+al11}. The inner mass distribution of (most of) these clusters seem correspondingly rather widely-distributed (see also Table \ref{table:sample}), which may account for their extensive lensing properties and large Einstein radii. In this regard, the combined analysis of lensing data together with X-ray measurements and observations of the SZ effect seems to be a very promising approach to constrain the intrinsic structure of the mass and gas distributions even in unrelaxed clusters \citep{def+al05,ser+al06,ser+al11,mor+al11}.

Unlike relaxed and highly-concentrated clusters which correspondingly have more mass in the middle boosting their critical area and Einstein radius, in higher-$z$, unrelaxed clusters, large Einstein radii may however form due to a widely-distributed inner mass distribution, so that the critical curves of the smaller substructures near the core merge together to form a much larger Einstein radius curve \citep{Tor+al04,Fed+al06,zit+al09}. If this is indeed the case, low concentrations and relatively shallow inner profiles should be expected for clusters as those in the $z>0.5$ MACS sample \citep{net+al07}. Although the results of our analysis may not be projected on relaxed lower-$z$ samples or unambiguously resolve the claimed discrepancy in the $c(M)$ relation for relaxed halos, they are of an additional high importance, since such information can help constrain and shed new light on the relaxation era of (massive) clusters, and more generally on the evolution of the large-scale structure. Our finding of low concentrations suggests, as previously implied \citep[e.g.,][]{net+al07,zit+al11b}, that high-$z$ unrelaxed clusters therefore form a \emph{different class} of prominent and impressive gravitational lenses.

Our analysis would benefit from either arc redshift confirmation or data for different scales, as those provided by weak lensing observations. Upcoming observations planned in the CLASH program \citep{Post+al11CLASH} are going to supply this information and will further establish the results of our work. However, combined strong plus weak lensing analyses may be dominated by the strong lensing part due to the lower systematics. Therefore, considering as a first step only strong lensing can be reasonable.

\section*{Acknowledgements}
We thank Tom Broadhurst for very useful discussions.


\setlength{\bibhang}{2.0em}

\end{document}